# *On Five Independent Phenomena Sharing a Common Cause*
## *by*
## *Roger Ellman*

Abstract

Over the past century a succession of five different independent astronomical phenomena have been discovered, each appearing to be the result of a common underlying cause that also produces an unaccounted-for acceleration that is: quite small, centrally directed in the system exhibiting each phenomenon, non-gravitational, distance independent, and apparently of a common magnitude.

The present paper analyzes the phenomena and proposes the underlying common cause, a common solution to the problem that they present.

Four of the phenomena, in the order of their discovery are:

1 – In 1933, the indication by galactic rotation curves that there is such an acceleration present and acting in galaxies but with no observable cause [hence the postulating of "Dark Matter"]. Here the acceleration is directed toward the galactic center, the dominant factor in the mechanics of galaxy rotation.

2 – In 1998, the Pioneer Anomaly in which the acceleration is directed toward the Sun, the dominant factor in the mechanics of the Pioneer spacecrafts' motion.

3 – In 2008, the Flybys Anomaly for which the acceleration is directed toward the center of the Earth, the dominant factor in the mechanics of the flyby motion [as presented in the paper following].

4 – Also in 2008, confirmed in 2010, the Dark Flow anomaly for which the acceleration is directed toward the central origin of the overall universe, the dominant factor in the mechanics of the overall universe, where the Big Bang and expansion began [as presented in the paper following].

The earliest of all is

5 – Hubble's discovery of Redshifts of the light from various distant astral bodies.

Roger Ellman,   The-Origin Foundation, Inc.
320 Gemma Circle, Santa Rosa, CA 95404, USA
RogerEllman@The-Origin.org
707-537-0257
http://www.The-Origin.org

# *On Four Independent Phenomena Sharing a Common Cause*
## *by*
## *Roger Ellman*

## *The Problem*

Four independent unrelated phenomena, none of which has an established explanation, have now been extensively observed and a large amount of data substantiating the phenomena have been developed. The phenomena are as follows.

- In 1933 F. Zwicky reported [1] that the rotational balance of gravitational central attraction and rotational centripetal force in galaxies appeared to be out of balance, that a small additional centrally directed acceleration of unknown source appeared to be needed and to be acting. Numerous galactic rotation curves confirm that there is such an anomalous acceleration present and necessary in all rotating galaxies.

- In 1998 the Pioneer Anomaly was first reported [2]. The anomaly is a small acceleration, centrally directed [toward the Sun], constant, distance independent, and of unknown cause, observed in the tracking of the Pioneer 10 and 11 spacecraft from launch until their near departure from the Solar System.

- In 2008 the Flybys Anomaly was first reported [3]. The anomaly is unaccounted for changes in spacecraft speed, both increases and decreases, for six different spacecraft involved in Earth flybys from December 8, 1990 to August 2, 2005.

- Also in 2008 a previously unknown large scale flow of galaxy clusters all in the same direction toward "the edge" of the observable universe, the Dark Flow anomaly was first reported [4]. The mysterious motion, originally noted in 2008 using the three-year WMAP survey, is now [2010] confirmed by a more comprehensive five-year study [7].

Analysis discloses that the first three have in common the same locally centrally directed, small acceleration that is non-gravitational, distance independent, constant, and unaccounted for. The fourth phenomenon is shown to be fully consistent with that same cause and explanation.

A fifth phenomenon and the earliest, astral Redshifts discovered by E. Hubble before the above, turns out to be primarily caused by the same underlying phenomenon causing the others.

## *The Centrally Directed Anomalous Acceleration in all Rotating Galaxies*

In general, galaxies are rotating systems, a balance of gravitational attraction $[G \cdot M \cdot m / R^2]$ and centripetal force $[m \cdot V^2 / R]$ maintaining the structure. A curve or plot of such rotational velocity, $V$, versus path radius, $R$, is termed a Rotation Curve.

When the central mass is far greater than the orbiting masses the dynamics are such that the orbital velocities are inversely proportional to the square root of the radial distance from the center mass $[V = (G \cdot M / R)^{1/2}]$, as for example in our solar system and as illustrated in Figure 1, below. Such rotational dynamics and rotation curves are referred to as Keplerian.

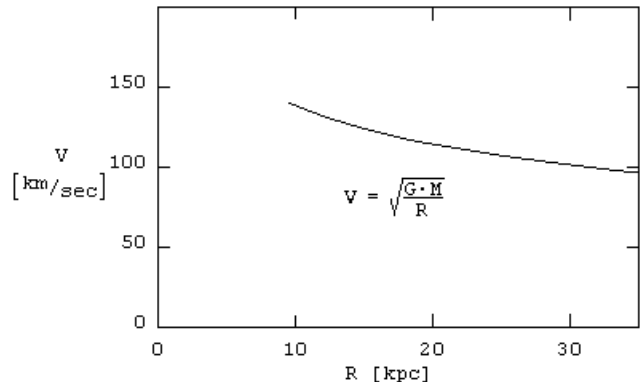

*Figure 1 - A Keplerian Rotation Curve*

In the case of a solid sphere of uniform density, $\rho$, throughout, all parts must move at rotational velocities directly proportional to radius as illustrated in Figure 2, below.

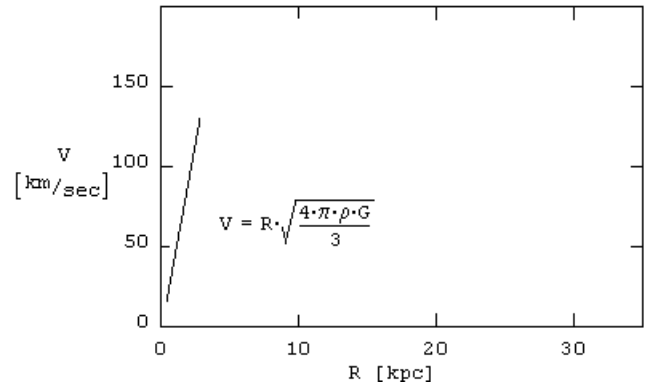

*Figure 2 - The Rotation Curve of a Solid Sphere of Uniform Density*

The form of galaxies as we are able to directly observe them is that of a fairly spherical star-dense central core and a transition from that to the much more extensive flat disk of a far smaller density of more widely dispersed stars. The portion of galactic rotation curves that pertains to the dense central core of the galaxy would be expected to exhibit approximately the same velocity-proportional-to-radius form as illustrated for a solid sphere in Figure 2, above. Likewise, the more dispersed flat disk, minor in mass compared to the dense central core, would be expected to exhibit the Keplerian form of Figure 1, above. The expected form of galactic rotation curves would be the two combined with a smooth transition between as Figure 3, below.

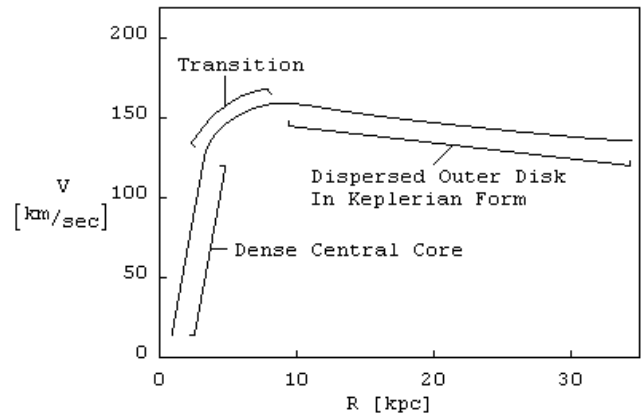

*Figure 3 - The Expected Form of Galactic Rotation Curves*

For galaxies that present themselves in an edge view of the thin disk not as their spiral or globular spread in space, it is possible to measure the rotational velocities and obtain a rotation curve. We see one end of the presented flat disk moving toward us relative to the center and the other end moving away. The rotational velocities are measured along the galactic diameter represented by our view of the disk by observing the variations in redshift, those variations being a Doppler effect. Galactic rotation curves so obtained do not exhibit the expected Keplerian form, an inverse square root of radius. Rather, they exhibit a flat form, that is, they exhibit rotational velocity independent of radius. The overall curve, after the portion pertaining to the dense central core of the galaxy, is a transition to a flat curve in the region corresponding to the spread-out galactic disk as in Figure 4, below.

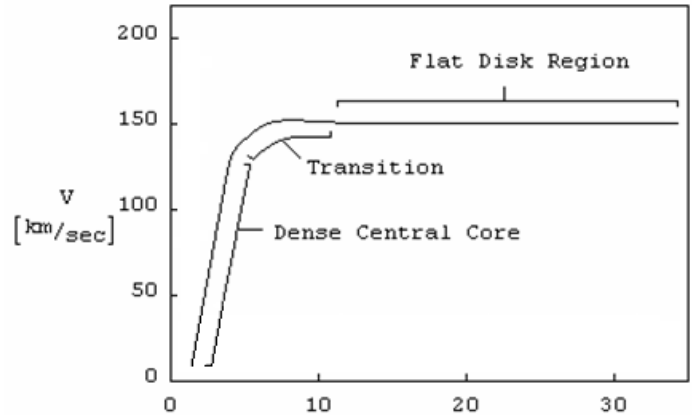

*Figure 4 - A Typical Galactic Rotation Curve as Observed*

Because the form of the flat portion of galactic rotation curves lies between the case of a dominant central mass, as in the Keplerian inverse square root of radius form [Figure 1], and the case of a uniformly dense mass, with its direct proportion to radius form [Figure 2], it has been inferred that matter that we have not observed must be present similarly distributed within the galaxy. That is, it is inferred that unobservable matter must be distributed in the galaxy in a manner that lies between the matter distribution of a dominant central mass [the Keplerian case] and that of a uniformly dense mass [the direct proportion to radius case] as a halo of "dark matter" which causes the rotation to take the form that the rotation curve exhibits. Thus arose the "dark matter" hypothesis.

No explanation has been offered for why the "dark matter", while performing a gravitational function in the galaxy nevertheless fails to be distributed in the same manner as the "visible matter" in a fairly spherical dense central core with a transition from that to a much more extensive flat disk which has a far smaller density of more widely dispersed stars

However, what the rotation curves demonstrate is not the existence of a hypothesized cause [dark matter]; <u>they only demonstrate the existence of an acceleration that is not accounted for</u>. That acceleration is identified as follows. A constant acceleration, $\Delta a_{Anomalous} = 8.7 \cdot 10^{-8}$ $cm/_{sec2} = a_A$, acting alone as a gravitational acceleration maintaining a mass in orbit, would produce a rotation curve as in Figure 5, below.

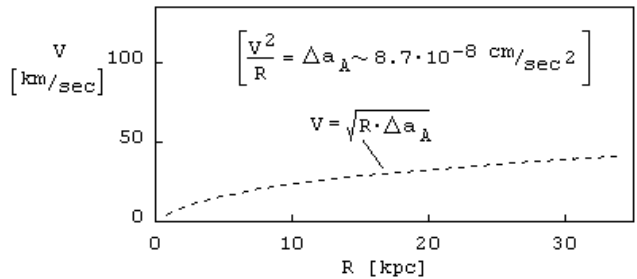

*Figure 5 - The Rotation Curve of $a_{Anomalous}$ Acting Alone*

That rotation curve is of the correct form and magnitude to convert a galactic rotation curve exhibiting a Keplerian form [as in Figure 1] to a flat one [as in Figure 4]. That is, the rotation curve of $a_{Anomalous}$ exhibits $V$ <u>directly</u> proportional to the square root of $R$ and the Keplerian rotation curve exhibits $V$ <u>inversely</u> proportional to the square root of $R$. The two effects tend to cancel and leave a flat rotation curve. With the naturally occurring typical rotation curve modified by the addition of $a_{Anomalous}$ the rotation curve becomes flat, as illustrated in Figure 6, below, by superimposing the curves.

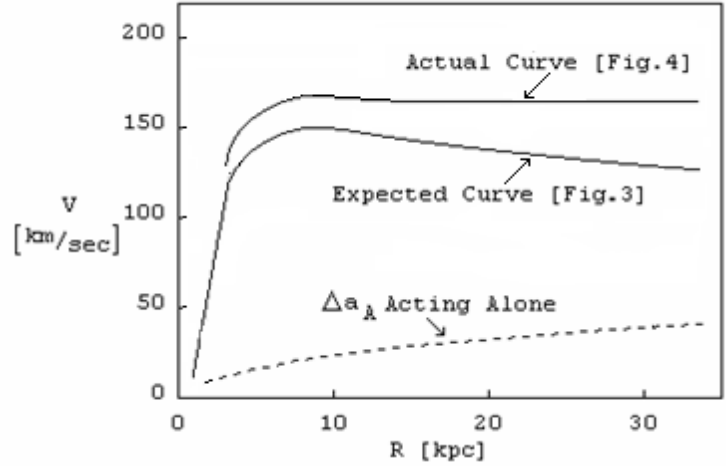

*Figure 6 - The Anomalous Acceleration, $a_{Anomalous}$, Acting Alone Superimposed on the Expected and Actual Rotation Curves [Figures 3 & 4]*

Of course, the rotational velocities corresponding to the components of the total acceleration cannot properly be added. Rather, the accelerations must be summed and the resulting rotational velocities then obtained as follows,

(5)  Total Acceleration = "natural acceleration" + $\Delta a_A$ = $a_{Anomalous}$]

$$\frac{V^2}{R} = \frac{G \cdot M}{R^2} + \Delta a_A$$

$$V = \left[\frac{G \cdot M}{R} + R \cdot \Delta a_A\right]^{1/2}$$

which produces the observed actual flat portion of the rotation curve in the region corresponding to where the "expected" form is Keplerian.

### *The Pioneer Anomaly*

The just preceding galactic rotation curve anomalous acceleration $\Delta a_{Anomalous} = 8.7 \cdot 10^{-8}$ cm/sec2 is identical in magnitude to the Pioneer Anomaly anomalous acceleration. The Pioneer Anomaly is a small acceleration of $8.7 \cdot 10^{-8}$ cm/sec2, centrally directed [toward the Sun], constant, distance independent, and of unknown cause. The evidence for it is abundant tracking data that have been reviewed and re-reviewed in search of error with the result that the effect is highly validated.

Since the original reporting of the Pioneer Anomaly in 1998 sources of systematic error external to the spacecraft [e.g. solar wind / radiation], internal to the spacecraft [e.g. gas leakage], and in the computational system [e.g. model accuracy / consistency] have all been thoroughly examined. All of those sources of error are either too small, not applicable, and / or act in the wrong direction to account for the phenomenon. The input of suggested sources of systematic error to those analyses has been not only from the research team of authors but from a number of other sources interested in the problem. The source area of systematics has been essentially exhausted.

The only difference between the Pioneer Anomaly acceleration and the galactic rotation curve anomalous acceleration is that in the Pioneer case the acceleration is directed toward the Sun, the dominant factor in the mechanics of the Pioneer spacecrafts' motion whereas the galactic rotation curve anomalous acceleration is directed toward the rotational center of the galaxy, the dominant factor in the mechanics of galaxy rotation.

## *The Flybys Anomaly*

In March 2008 anomalous behavior in spacecraft flybys of Earth was reported in Physical Review Letters, Volume 100, Issue 9, March 7, 2008, in an article entitled "Anomalous Orbital-Energy Changes Observed during Spacecraft Flybys of Earth"[1].

The data indicate unaccounted for changes in spacecraft speed, both increases and decreases, for six different spacecraft involved in Earth flybys from December 8, 1990 to August 2, 2005. These anomalous energy changes are a function of the incoming and outgoing geocentric latitudes of the asymptotic spacecraft velocity vectors and further indicate that a latitude symmetric flyby does not exhibit the anomalous speed change. The article states that, "All … potential sources of systematic error …. [have been] modeled. None can account for the observed anomalies…. "Like the Pioneer anomaly … the Earth flybys anomaly is a real effect …. Its source is unknown."

A phenomenon like that involved in galactic rotation curves and in the Pioneer Anomaly would account for the highly varied occurrences of the flyby anomaly: a small acceleration [in addition to that of natural gravitation], centrally directed and independent of distance; that is a modest and otherwise unknown acceleration directed toward the core center of the Earth, the principle body involved, the dominant factor in the mechanics of the flyby.

To observe the relation to the Flybys Anomaly of an otherwise unknown or un-detected anomalous, centrally directed, distance independent acceleration the first step is to consider a simple spacecraft pass of Earth where the pass is all at zero latitude as shown in Figure 7, on the following page. In the vectors analysis part of the figures $A$ is the full anomalous acceleration, $C$ is its component parallel to the direction of motion of the satellite, and $\theta$ is the angle between the direction of action of those two.

When the spacecraft is at a great distance out from Earth the spacecraft's motion is close to being directed toward the center of the Earth but not exactly so. A centrally directed acceleration there analyzed into components parallel and perpendicular to the spacecraft's motion would show most of the centrally directed acceleration acting to increase the spacecraft's speed.

As the spacecraft travels nearer to Earth that component parallel to its motion decreases, becoming zero at the closest approach to Earth. From that point on the parallel component acts in the opposite direction on the spacecraft, that is its effect is to decelerate the spacecraft not accelerate it. Ultimately the anomalous acceleration and anomalous deceleration experienced by the spacecraft become equal and cancel each other out leaving as the only flyby effect the gravitational boost, due to another effect, that is the overall purpose of the flyby.

Of the full centrally directed acceleration, $A$, the component, $C$, parallel to the path of the flyby in this case is

*(1)*  `C = A·Cos[θ]`

which is apparent if the flyby path is a straight line. However, the actual flyby path is somewhat curved by the Earth's gravitation. But, the anomalous acceleration is always centrally directed toward the core of the Earth so that $C$ is nevertheless as stated.]

a. <u>Polar View - Flyby</u>

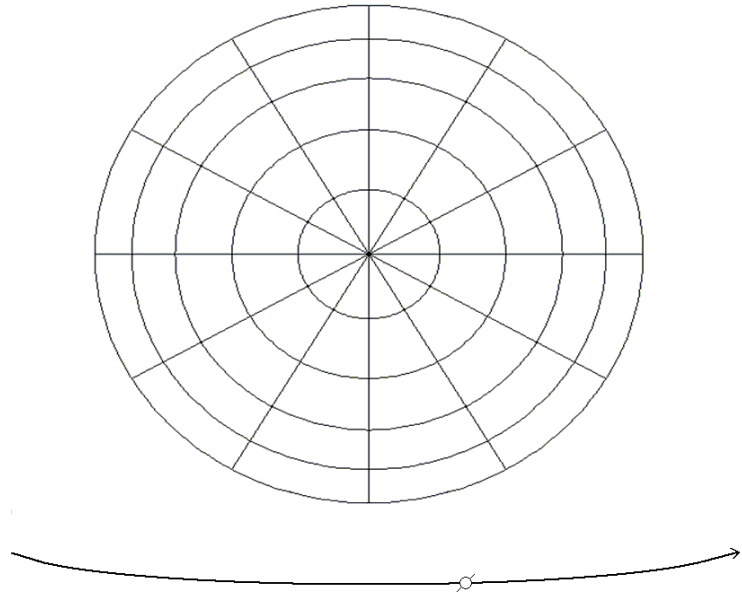

b. <u>Polar View - Anomalous Acceleration Vectors</u>

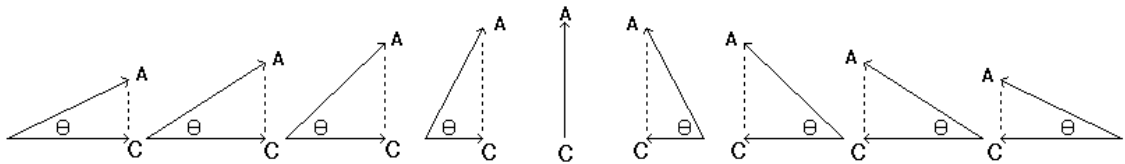

c. <u>Equatorial View</u>

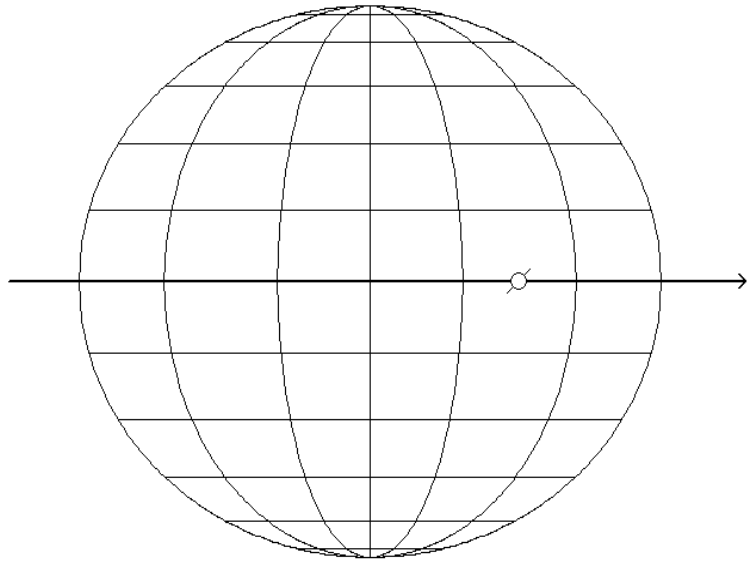

*Figure 7*
*A Zero Latitude Pass*

Equation *(1)* is valid when the flyby pass is solely at zero latitude. However, if other than zero the latitude of the flyby pass has a significant effect on the magnitude of `C`, the component of the overall centrally directed acceleration parallel to the spacecraft flight path. As latitude increases the magnitude of `C`, decreases. That is most easily visualized by imagining the flyby over the geographic north pole at `90°` north latitude. There the centrally directed acceleration toward the center of the Earth has no component parallel to the flight path.

Therefore, for flyby paths at other than zero latitude the effective value of `A` is `A(λ)` a function of latitude, $\lambda$, as equation *(2)*

*(2)* `A = A(λ) = A·Cos[λ]`

so that equation *(1)* then becomes equation *(3)* the full expression for the extent to which the centrally directed anomalous acceleration actually accelerates or decelerates the spacecraft.

(3) `C = A·Cos[λ]·Cos[θ]`

The gross effect of latitude can be evaluated by examining three cases:

  A - The flyby path is symmetrical relative to the equator so that the latitude effect in the first half of the flyby, `θ = 0°` to `90°`, is exactly offset or balanced by the second half of the flyby, `θ = 90°` to `180°`. This case is essentially the same as presented in Figure 7, above.

  B - The flyby path starts at low latitude and finishes at high latitude, Figure 8 on the following page.

  C - The flyby path starts at high latitude and finishes at low latitude, Figure 9 on the second following page.

Per the equations and Figure 7 in the first half of the flight path the effect of the anomalous, centrally directed acceleration is to increase the speed of the spacecraft whereas the effect in the second half of the flight path is to decrease the spacecraft's speed. By its definition Case A produces no net anomalous acceleration or deceleration of the spacecraft because the first and second halves of the flight path balance and offset each other.

In Case B, the first half, i.e. the acceleration half, of the flight path is at low latitude where the latitude effect only modestly reduces the anomalous acceleration magnitude. But for that case and path the second half, i.e. the deceleration half, of the flight path is at a high latitude where the latitude effect greatly reduces the anomalous acceleration magnitude. The net effect is a relatively large acceleration followed by a lesser deceleration for a net increase in the spacecraft's speed.

In Case C, the effect is just the reverse of that in Case B; the first, i.e. the acceleration, half of the flight path is at high latitude where the effect of the latitude greatly reduces the anomalous acceleration magnitude. But for that case and path the second, i.e. the deceleration, half of the flight path is at a low latitude where the effect of the latitude only modestly reduces the anomalous acceleration magnitude. The net effect is a relatively small acceleration followed by a greater deceleration for a net decrease in the spacecraft's speed.

Therefore, depending on the specific flight path of the spacecraft's flyby pass of Earth the spacecraft may experience an overall net anomalous acceleration or a net anomalous deceleration, those in various amounts depending on the specific encounter and the latitudes involved, and zero net modification if the path is perfectly latitude symmetrical.

a. Equatorial View - Flyby

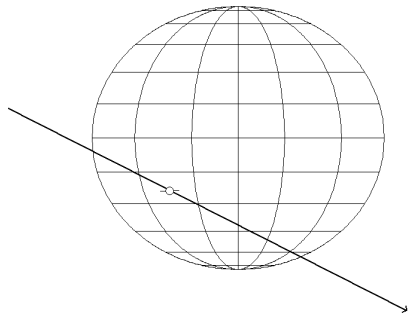

b. Equatorial View - Flyby, Rotated

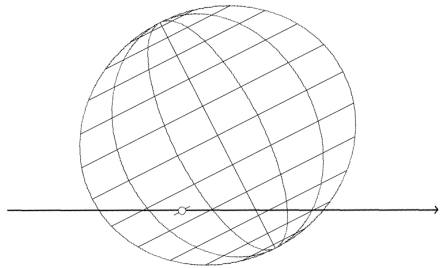

c. Anomalous Acceleration Vectors

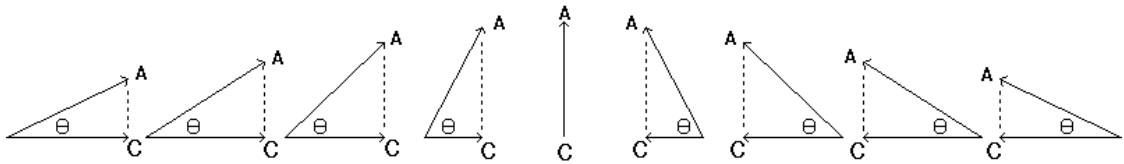

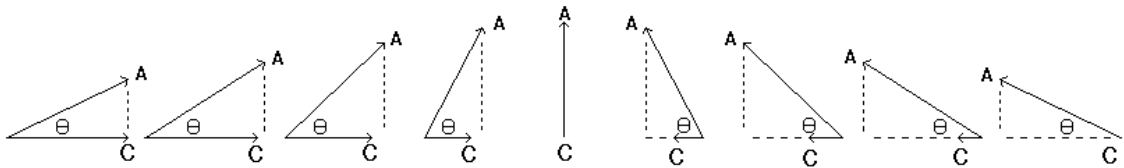

The result in this case is a net acceleration
[to the right in the diagrams].

*Figure 8*
*A Pass at Increasing Latitude*

a. <u>Equatorial View - Flyby</u>

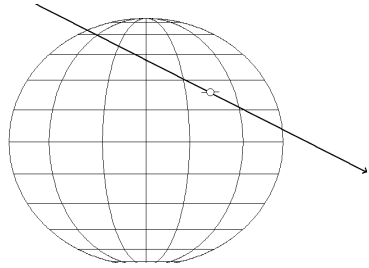

b. <u>Equatorial View - Flyby, Rotated</u>

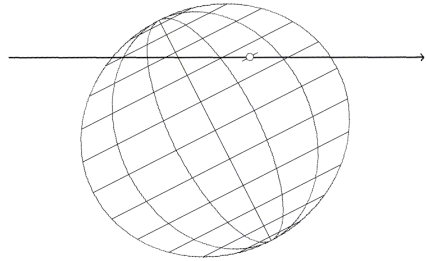

c. <u>Anomalous Acceleration Vectors</u>

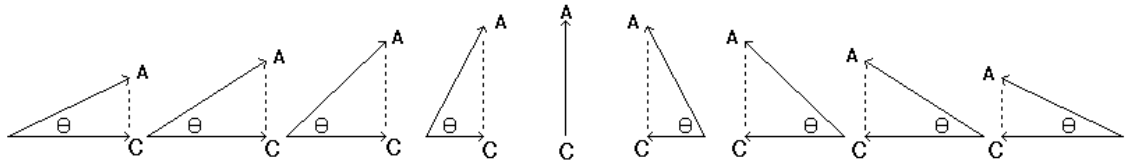

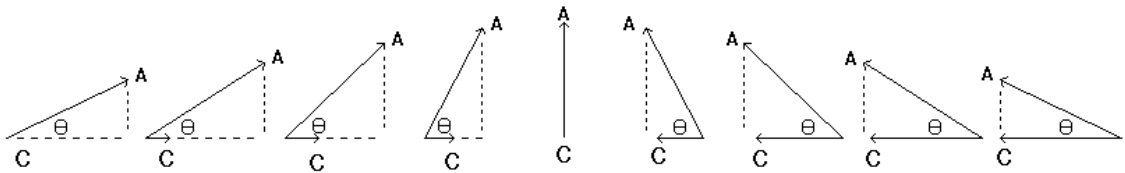

The result in this case is a net deceleration, that is an acceleration toward the left in the diagrams, against the direction of the flyby.

*Figure 9*
*A Pass at Decreasing Latitude*

## The "Dark Flow" Anomaly

Thus there are small, centrally directed, distance independent, non-gravitational, same, anomalous accelerations appearing as a near Earth effect [the Flybys Anomaly], a Solar effect [the Pioneer Anomaly], and a galactic effect [galactic rotation curves]. It can only be concluded that the same effect must appear relative to every planet [and every planet's moons], every sun [star], every galaxy and every group of galaxies.

And such a small, centrally directed, distance independent, non-gravitational, same, anomalous acceleration could be expected to appear for the universe overall, directed toward the center of the universe, the location of the origin, where it all began.

The universe began with the "Big Bang", an immense explosion radially outward in all directions, largely spherically symmetrically, from an original source "singularity".

We, residing on planet Earth, of star Sol, in one of several branches of spiral galaxy Milky Way, are located off some significant distance in "our general direction" from and relative to the location of the original singularity.

We can "see" or detect a large number of neighbor galaxies, distant and near, whose components similarly proceeded outward from that "Big Bang" in directions slightly or significantly other than our particular direction.

But, there is a further mass of stellar bodies that proceeded outward from the "Big Bang" in directions away from us. What we can detect is only well less than half the total product of the "Big Bang".

The original location of the singularity, the origin, lies essentially at the center of the largely spherical volume of the source's product, the expanding universe. And the universe that we "see" lies largely to one side of that origin's location

To the above list of three effects caused by the systematic contraction of the universe can now be added the "Dark Flow" as originally reported in 2008 and recently further analyzed in terms of extensive new data as reported in NASA Goddard Release No.: 10-023 and in A. Kashlinsky, F. Atrio-Barandela, H. Ebeling, A. Edge, and D. Kocevski. *A New Measurement of the Bulk Flow of X-Ray Luminous Clusters of Galaxies*. [7].

> Distant galaxy clusters mysteriously stream at a million miles per hour towards a single point in the sky, separate from the expansion of the universe, along a path roughly centered on the southern constellations Centaurus and Hydra. A new study led by Alexander Kashlinsky at NASA's Goddard Space Flight Center in Greenbelt, Md., tracks this collective motion -- dubbed the "dark flow"....
>
> The clusters appear to be moving along a line extending from our solar system toward Centaurus / Hydra ... away from Earth. The distribution of matter in the observed universe cannot account for it. Its existence suggests that some structure beyond the visible universe -- outside our "horizon" -- is pulling on matter in our vicinity.

This is indication of the overall universe's experiencing an anomalous centrally directed acceleration accelerating all the matter of the universe gradually back toward the location of its origin [as described above]. This "Dark Flow" is part of that centrally directed acceleration toward the location of the origin of the universe, a location at or just beyond the "edge" of the universe "see-able" by us.

A "map" of the universe that we "see" would look somewhat as in Figure 10, below, where the regions of galaxies studied involved in the "dark flow" are indicated in the large colored areas in red, yellow, green and blue.

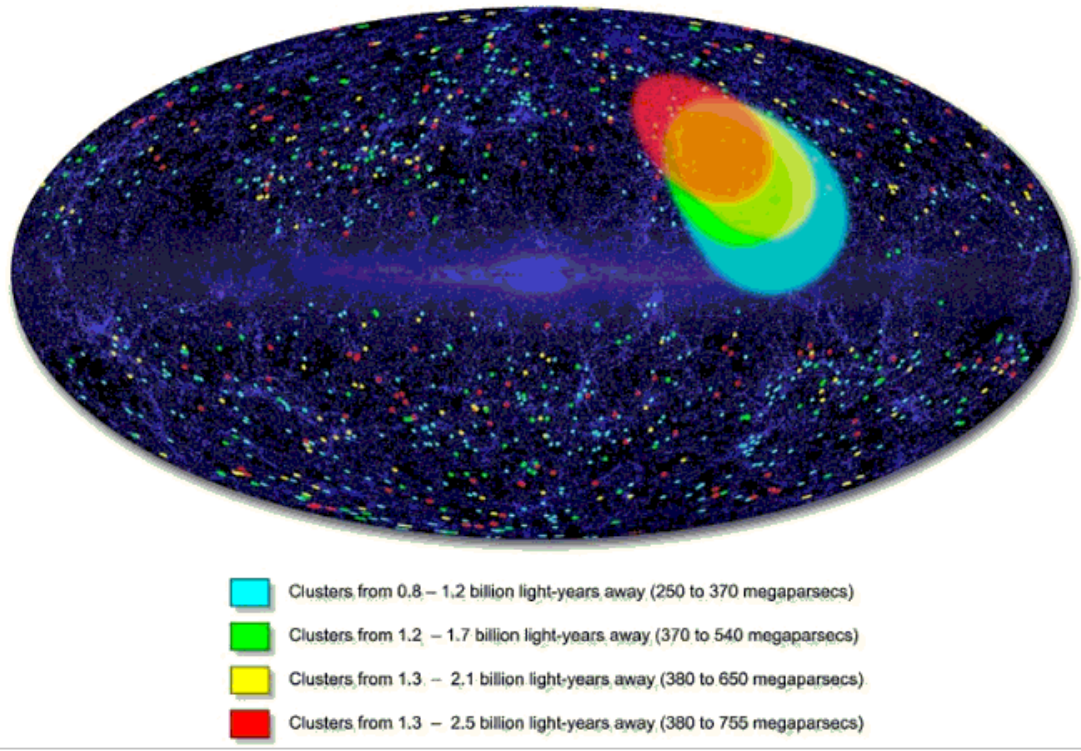

*Figure 10*
*A Map of That Part of the Universe Observable to Us*

## *A General Anomalous Acceleration Throughout the Universe*

Thus there are small, centrally directed, distance independent, non-gravitational, same, anomalous accelerations appearing as a near Earth effect [the Flybys Anomaly], a Solar effect [the Pioneer Anomaly], a galactic effect [galactic rotation curves], and a Universe effect [the Dark Flow]. It can only be concluded that the same effect must appear relative to every planet [and every planet's moons], every sun [star], every galaxy and group of galaxies, and the universe overall. In other words as a general cosmic effect.

What could produce such a phenomenon ? What would cause there to be a universe-wide occurrence of such same accelerations ?

Taken together, planet relative, star relative, galaxy relative, universe relative, they collectively are a systematic contraction, a gradual reduction in the length component of every physical quantity in the universe. A general universal decay.

In material reality such decays are exponential. There are myriad examples of such, for example: radioactive decay, the decay of electrical transients in circuits involving inductance and capacitance, the decay of motion transients in mechanical systems involving mass and spring, the amplitude decay in a rung bell or a plucked string, etc. It is not unreasonable that a universe that began with an explosive "bang" follow that with a gradual exponential decay.

Such a decay of the overall universe was predicted and analyzed in detail in 1998, before the appearance of all except the earliest of the foregoing various anomalies, in the book *The Origin and Its Meaning*, Section 21 [5].

The Universal Exponential Decay is an exponential decay of the length dimensional aspect of all quantities in the universe. It involves the fundamental constants ($c$, $q$, $G$, $h$, etc.)

and decay of any of those must be dimensionally consistent with the decay of the others. The dimension that is decaying is length, the *[L]* dimension in the dimensions of, for example: h, *[M·L² /T]*; c, *[L/T]*; and G, *[L³/M·T2]*. The time constant of the decay is about τ = 3.57532·10¹⁷ sec (≈ 11.3373·10⁹ years).

Objections that such an effect would conflict with the known planetary system performance per the highly accurate planetary ephemeris are a mistaken interpretation of the situation. Consider a planet in circular orbit around a sun as in Figure 11, below.

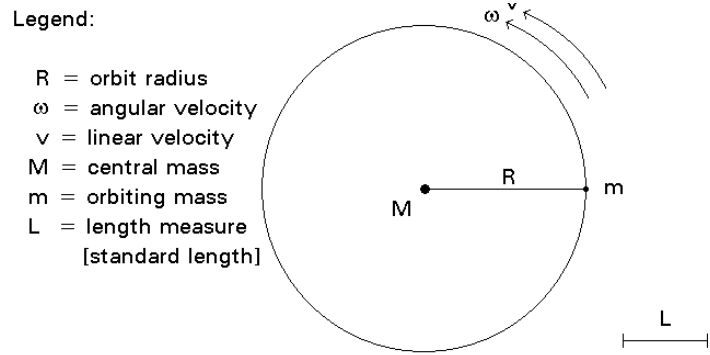

*Figure 11*

The relationship governing the motion is, of course, equation *(4)*, below

*(4)* Centripetal Acceleration Required = Gravitational Attraction Acceleration

$$V^2/R \text{ (or) } R \cdot \omega^2 = G \cdot M / R^2$$

Now, let the length dimensional aspect [with the dimensions of all quantities expressed in the fundamental dimensions of mechanics, *[L], [M],* and *[T]*] of all quantities decay, becoming gradually smaller with time. That is, let all lengths, *[L]*, decrease by being multiplied by the decay function, *D(t)*, per equation *(3)*, below. [For the present purpose the form of the decay function is irrelevant except that it must be a function of time. The decaying exponential is used because it is common in nature and is a complicated case.]

*(5)* $D(t) \equiv \varepsilon^{-[t/\tau]}$, where τ is the time constant of the decay

Then the quantities involved in equation *(4)* all change to as follows.

*(6)* <u>The Orbital Radius, R, [dimension = **L**]</u>

R becomes $R(t) = R(t=0) \cdot \varepsilon^{-[t/\tau]}$

<u>The Gravitational Constant [dimensions = **L³**/M·T 2]</u>

G becomes $G(t) = G(t=0) \cdot \{\varepsilon^{-[t/\tau]}\}^3$

<u>The Centripetal Acceleration Required [dimensions = **L**/T2]</u>

$R \cdot \omega^2$ becomes $R(t) \cdot \omega^2 = [R(t=0) \cdot \varepsilon^{-[t/\tau]}] \cdot \omega^2$

$$= [R(t=0) \cdot \omega^2] \cdot \varepsilon^{-[t/\tau]}$$

or

$$\frac{V^2}{R} \text{ becomes } \frac{[V(t)]^2}{R(t)} = \frac{[V(t=0) \cdot \varepsilon^{-[t/\tau]}]^2}{[R(t=0) \cdot \varepsilon^{-[t/\tau]}]}$$

$$= \frac{[V(t=0)]^2}{R(t=0)} \cdot \varepsilon^{-[t/\tau]}$$

*The Gravitational Attraction Acceleration*  [dimensions = $L/T^2$]
[and where the G dimensions = $L^3/M \cdot T^2$]

$$\frac{G \cdot M}{R^2} \text{ becomes } \frac{G(t) \cdot M}{[R(t)]^2} = \frac{[G(t=0) \cdot \{\varepsilon^{-[t/\tau]}\}^3] \cdot M}{[R(t=0) \cdot \varepsilon^{-[t/\tau]}]^2}$$

$$= \frac{G(t=0) \cdot M}{[R(t=0)]^2} \cdot \varepsilon^{-[t/\tau]}$$

The overall net effect is: $R$ decreases, the required centripetal acceleration decreases in proportion, the gravitational attraction likewise decreases in proportion, and $\omega$ is unchanged.

Furthermore, we observers, using our measuring standard ruler, length $L$ of the above Figure 11, would never detect any of the decay because our standard length would also be decaying at exactly the same rate, in the same proportion.

The point of this obvious mathematics / physics exercise is that a universal decay of the length aspect of all material reality would not conflict with the planetary ephemeris and would not even be detectable at all except in unusual circumstances such as the Pioneer and Flyby anomalies and the evidence of galactic rotation curves; nor would it interfere with the relative values of the fundamental constants and their interactions in physical laws.

Returning to the orbiting body of Figure 11, reproduced as Figure 12 below, the figure's annotations slightly modified, the development of the anomalous acceleration is very direct.

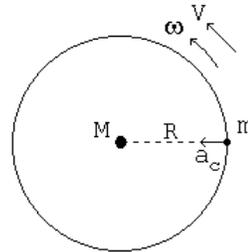

Legend:
- R = orbit radius
- ω = angular velocity
- V = linear velocity
- M = central mass
- m = orbiting mass
- $a_c$ = centripetal acceleration
  = Newtonian gravitation + anomalous acceleration $a_p$

*Figure 12*

The Newtonian component of the centripetal acceleration is only sufficient to maintain the orbit, to keep $R$ constant, to prevent its increasing. For the orbiting body, $m$, to gradually approach the central mass, $M$, that is for $R$ to decrease, additional inward acceleration is required.

That inward acceleration is the anomalous acceleration appearing as a near Earth effect [the Flybys Anomaly], a Solar effect [the Pioneer Anomaly], and a galactic effect [galactic rotation curves]. It is an unavoidable concomitant effect of the contraction of the length dimension $[L]$ of $R$ in the above example and of the systematic contraction, the gradual reduction in the length component, of every physical quantity in the universe, of all material reality.

## *Summary Conclusions*

1 – All four effects: the Galactic Rotation Curves Anomaly, the Pioneer Anomaly, the Flybys Anomaly, and the Dark Flow involve the same common action, a small, centrally directed, non-gravitational, distance independent acceleration, apparently the same common acceleration $\Delta a_{Anomalous} = 8.7 \cdot 10^{-8}$ cm/sec$^2$.

2 – The occurrence of such an acceleration apparently universe-wide is indicative of an on-going general contraction of the length aspect of all material reality including the length dimensional aspect of all fundamental constants.

Details on the universal contraction, or decay -- its cause, origin and characteristics are too lengthy for this report and are provided in full in reference [5].

## The Universal Decay

The Universal Decay causes the speed of light now to be a smaller, decayed value relative to light speed earlier. Thus in general the speed of light is $c(t) = C_0 \cdot \varepsilon^{-t/\tau}$. [$C_0$ is the original speed of light at the instant of the "Big Bang" and $t$ is time since the "Big Bang"].

The speed of light is now decaying from its present value as we know it, $c$ or $c_{now}$, as $c(t) = c_{now} \cdot \varepsilon^{-t/\tau}$. Therefore the rate of change of the speed of light now is as follows.

$$(7) \quad \frac{d[c(t)]}{dt} = -\frac{c_{now}}{\tau} = -\frac{2.99792 \cdot 10^{10}}{3.57532 \cdot 10^{17}} = -8.38504 \cdot 10^{-8} \text{ cm/s}^2$$

compared to the Pioneer Anomaly $= -(8.74 \pm 1.33) \cdot 10^{-8}$ cm/s$^2$

That rate of change of the speed of light is due to the rate of change of its length dimensional aspect and, therefore, is the at present rate of change of all length dimensional aspects. It is the rate of the universal contraction, the un-accounted for centrally directed acceleration demonstrated in galactic rotation curves, the Flybys Anomaly and the Dark Flow Anomaly.

Because the decay time constant is so large the at-present rate appears to us to be constant.

Because everything including our instrumentation, our measurement standards, our atoms and ourselves are all experiencing the same decay, the decay is unnoticeable to us and is generally undetectable by us except for unusual circumstances such as the anomalies presented above.

## Validating The Universal Decay

Because the speed of light is decaying, light emitted long ago is faster than our present contemporary light, which causes the ancient light to appear to us to have a longer wavelength, that is, to be Redshifted. [Some of Redshifts, but not more than a minor portion, is due to the Doppler Effect of the astral sources' outward velocities.]

Aside from observation of Redshifts, <u>each observation of which is actually an observation of the universal decay</u>, there are two other specific experimental observations that can be conducted to verify the Universal Decay and the value of its decay time constant.

- It can be tested that the speed of the light from distant astral sources is larger than our contemporary light speed. The earlier procedure of Michaelson or Pease and Pearson using the Foucault method is now superseded by the modern procedure, which is to modulate the light beam and use that modulation to measure the time required for the light to traverse a known distance.

- It can be tested that the Planck Constant of the light from distant astral sources is larger than our contemporary Planck Constant, $h$, using the photoelectric effect. Measuring the retarding potential that reduces the photoelectric current to zero, for light spectrally selected of a specific frequency, plots [for a set of different frequencies] as diagonal straight lines whose slope is the Planck Constant of that light.

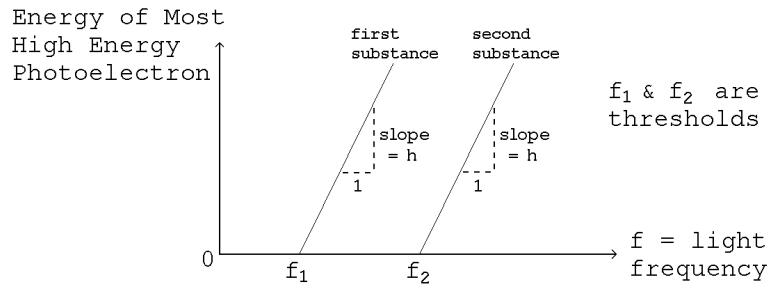

*Figure 13*